\documentclass[11pt]{article}
\usepackage{scicite}
\usepackage{times}
\usepackage{indentfirst}
\usepackage{diagbox}
\usepackage{makecell}
\usepackage{multirow}
\usepackage{enumerate}
\usepackage{amsmath}
\usepackage[most]{tcolorbox}
\usepackage{eqnarray,amsmath}

\newtcolorbox{myframe}[1][]{
  enhanced,
  arc=0pt,
  outer arc=0pt,
  colback=white,
  boxrule=0.8pt,
  #1
}

\newenvironment{sciabstract}{%
\begin{quote}}
{\end{quote}}

\newcounter{lastnote}

\title{Predict Forex Trend via Convolutional Neural Networks} 
\author
{Yun-Cheng Tsai,$^{1}$\footnote{The author was supported in part by the Ministry of Science and Technology of Taiwan under grant 106-3114-E-001-005. To whom correspondence should be addressed; E-mail: pecutsai@ntu.edu.tw}\; Jun-Hao Chen,$^{2}$\; Jun-Jie Wang$^{3}$ 
\\
\\
\normalsize{$^{1}$Center for General Education}\\
\normalsize{$^{2,3}$Department of Computer Science and Information Engineering}\\
\normalsize{$^{1,2}$National Taiwan University, Taipei 10617, Taiwan}\\
\normalsize{$^{3}$National Taipei University, New Taipei City 23741, Taiwan}
}

\date{}

\begin{document} 

\baselineskip15pt

\maketitle 

\begin{sciabstract}
Deep learning is an effective approach to solving image recognition problems. People draw intuitive conclusions from trading charts; this study uses the characteristics of deep learning to train computers in imitating this kind of intuition in the context of trading charts.

The three steps involved are as follows:
\begin{enumerate}
\item Before training, we pre-process the input data from quantitative data to images.
\item We use a convolutional neural network (CNN), a type of deep learning, to train our trading model.
\item We evaluate the model's performance in terms of the accuracy of classification.
\end{enumerate}
A trading model is obtained with this approach to help devise trading strategies. The main application is designed to help clients automatically obtain personalized trading strategies.

\end{sciabstract}

\section{Introduction}
Human beings are visual animals; the eyes are the most compact structure of all the sensory organs, and the visual intelligence of the human brain is rich in content. Exercise, behaviour, and thinking activities use visual sensory data as their greatest source of information. The more flexible and talented we become, the more we rely on visual intelligence.

What general business and decision makers desire after analysis is not the data itself, but the value. Therefore, it is important that data analyses be intuitive; in this way, the visualization of financial data can be more easily accepted: they can "see the story" and thus interpret the data more easily.

Although visualization analysis can benefit decision makers, many traditional statistical or machine learning methods for predicting currency movements use quantitative models. These methods do not consider visualization. We attempt to make good use of the advantages of visualization and comprehensively enhance the efficiency of intelligence analysis. For example, most traders use charts to analyse and predict currency movement trends, which carry obvious economic benefits. However, in this visualization, analysis is artificial. We intend to teach machines to achieve visualization like a human brain; we then hope to use the machine to visually analyse huge financial data.

Convolutional neural networks (CNNs) are widely used in pattern and image recognition problems. In these applications, the best possible correction detection rates (CDRs) have been achieved using CNNs. For example, CNNs have achieved a CDR of $99.77\%$ using the Modified National Institute of Standards and Technology (MNIST) database of handwritten digits, a CDR of $97.47\%$ with the New York University Object Recognition Benchmark (NORB) dataset of 3D objects, and a CDR of $97.6\%$ on over $5600$ images of more than $10$ objects. CNNs not only give the best performance compared to other detection algorithms but also outperform humans in cases such as classifying objects into fine-grained categories such as the particular breed of dogs or species of birds.

The two main reasons for choosing a CNN model to predict currency movements are as follows:
\begin{enumerate}
\item CNN models are good at detecting patterns in images such as lines. We expect that this property can also be used to detect the trend of trading charts.  
\item CNNs can detect relationships among images that humans cannot find easily; the structure of neural networks can help detect complicated relationships among features.
\end{enumerate}
CNN is a graph-based model, which is different from quantitative models. People do not need to consider all possible features that affect currency movements using quantitative models alone.

Compared to a quantitative model, a CNN model contains many filters that are similar to the eyes of a human being and can extract the features of images. As the convolution layer goes deeper, a CNN model can also extract more detailed features from the image, just like human visualization.

Predicting currency movement trends is a time-series problem. Many people look for the Holy Grail of prediction, which in fact does not exist. We cannot predict the future in the real world; however, we can define the small world to evaluate our prediction approach. In order to realize the idea, we use a geometric Brownian motion (GBM) to model the currency movements. We believe that these prices follow, at least approximately, as a subset of real-world rules that we can derive from the historical data and our knowledge of prices.

The three steps involved are as follows:
\begin{enumerate}
\item Before training, pre-process the training data from quantitative data to images. Our input images include price, Moving Average 5, Moving Average 10, and Moving Average 20 information.
\item Use a CNN to train our trading models.
\item Evaluate the models in terms of the accuracy of classification.
\end{enumerate}
When we control our small world, we use the CNN model to classify the weekly currency movements by separating price series into three groups: rising trend, down-trend, and non-movement groups. The remainder of this paper is organized as follows. A review of the literature is given in the next section. In Section 3, we present our methodology. Then, a description of the empirical data employed in our study is provided in Section 4. Section 5 presents the conclusion of our study. 

\section{Preliminary}
We use a graph-based model to train a predictive model, rather than using common quantitative methods such as recurrent neural networks (RNNs). In other words, we want to model the thoughts of people rather than the rule-based decisions, which can be clearly stated by the people.

Research on using CNNs to predict financial asset prices is limited; most researchers prefer the quantitative-based models. However, there are still some researchers attempting to study it.

Di Persio et al.~\cite{diartificial} tried to compare different artificial neural network (ANN) approaches to predict stock market indices in classification-based models. They compared three common neural network models, namely multilayer perceptron (MLP), CNN, and long short-term memory (LSTM). They found that a novel architecture based on a combination of wavelets and CNNs reaches an $83\%$ accuracy rate on foreign exchange rates, outperforming the RNNs by $4\%$.

Distinct from our work, Di Persio et al.~\cite{diartificial} designed their CNN architecture by using a 1-dimensional convolution layer and a 1-dimensional pooling layer. The 1-dimensional convolution layer considers only the price data, which means this convolution layer still captures the quantitative information.

Similar to our work, Ashwin Siripurapu used convolution networks to predict movements in stock prices with a series of time-series pixel images. The input images to the model are the graphs of high and low prices for a 30-min window of time. The input graphs to the model are saved in an RGB color space to highlight the different lines of the stock prices.

Siripurapu used three kinds of input images. For the first input, he used only the high and low prices, and for the second one, he added the volume data together with the high and low prices. For the third one, he used the correlation feature representation of the top ten companies share of the Standard and Poor's 500 index basket. In the experiment, Siripurapu used two different architectures of conventional networks, called full and reduced models. The full model had five pairs of convolution-ReLU-pooling layers and was further connected to a fully connected layer. The reduced model reduced the pooling layers in the first two pairs. Although the performance does not exceed 0 for an out-of-sample R square, it still gives us many ideas for using pixel images as the input data to a CNN model.

People like to think intuitively when viewing trading charts; many of them cannot clearly explain how to make their decisions and how to achieve better performance. We focus on this direction by using pixel images as inputs to enable the computer refine the features from it. However, beyond learning, we want to teach the computer to simulate, and thus predict the behaviour of people as they trade on the trading charts; that is, make a model which can learn the trading strategies of the people.

Define $S_t$ as the price of the financial asset at time $t$. The risk-neutralized version of stock price's log-normal diffusion process is
\begin{align}\label{GBMSDE}
dS_t=r\,S_t\,dt+\sigma\,S_t\,dW_t,
\end{align}
where $r$ is the risk-free rate, $\sigma$ is the constant volatility price process of the financial asset, and the random variable $W_t$ is a standard Brownian motion~\cite{browne1995optimal}.
$S_t$ is said to follow a geometric Brownian motion (GBM) process because it satisfies the above stochastic differential equation (SDE).
For an initial value $S_0$, the equation (\ref{GBMSDE}), has the analytic solution:
\[
S_t=S_0\exp{\left(r-\frac{\sigma^2}{2}\right)t+\sigma\,W_t}.
\]
From equation (\ref{GBMSDE}), it has the following discrete solution~\cite{shrevestochastic}:
\begin{align}\label{GBM}
X_t=X_{t-1}+\left(r-\frac{\sigma^2}{2}\right)\Delta t+\sigma\sqrt{\Delta t}\,B_{t},
\end{align}
where $X_t\equiv \ln(S_t)$ is the log-price, $\Delta t\equiv T/n$ is the length of a time step in the time interval $[0,T]$ divided into $n$ subintervals, $B_t\sim N(0,1)$ is i.i.d. normal random variable and $\sigma$ is the annualized constant volatility.

The CNN is one of the best graph-based models in recent years. Many new architectures of CNNs constantly appeared very fast, but the most original architecture was proposed by K. Fukushima in 1980. K. Fukushima proposed a model called Neocognitron, which is generally seen as the model that inspires the CNN on the computation side~\cite{fukushima1982neocognitron}.

Neocognitron is a neural network designed to simulate the human visual cortex. It consists of two types of layers, called feature extractor layers and structured connection layers. The feature extractor layers, also called S-layers, simulate the cell in the primary visual cortex, helping human beings to perform feature extraction. The structured connection layers, also called C-layers, simulate the complex cell in the higher pathway of the visual cortex, providing the model with its shifted invariant property.

Inspired by the Neocognitron and the concept of back propagation, the most generally classic modern CNN, LeNet, was proposed by LeCun et al. in 1990. The potential of the modern convolution architecture can be seen in LeNet (1990), consisting of a convolution layer, a subsampling layer, and a full connection layer~\cite{wang2017origin}.

As the concept of rectified linear unit (ReLU) and drop out were presented in recent years, a new convolution-based model, AlexNet, proposed by Hinton and Alex Krizhevsky, appeared and beat the previous champion of the ImageNet Challenge, with over 15M labelled high resolution images and roughly 22,000 categories. There are three main differences between LeNet and AlexNet:
\begin{enumerate}
\item The ReLU is used as the activation function in AlexNet. It introduces a non-linearity transform after convolution, which helps the computer to simulate human vision more accurately. The ReLU is also a non-saturating activation function and is several times faster than tanh and sigmoid units in computation.
\item A new regularization technique called drop-out was introduced to AlexNet to avoid over-fitting with much less computation. The drop-out technique randomly drops some neurons with a particular probability, and the dropped neurons are not involved in forward and backward computation.
\item Owing to the technological progress in recent years, AlexNet was supported by a more efficient GPU than LeNet (1980). This means that a larger dataset and more epochs can be trailed in the training process.
\end{enumerate}
With the success of AlexNet, many researchers have been motivated to participate in this kind of research, inventing architectures with deeper structures and modified convolution such as VGG and GoogleNet. These developments continually improve CNNs in the field of computer vision.

The two most important components of CNNs are the convolution layer and pooling layer. The convolution layer implements the convolution operation, which extracts image features by computing the inner product of an input image matrix and a kernel matrix; the number of channels of the input image and kernel matrix must be the same. For example, if the input image is an RGB color space, then the depth of the kernel matrix must be three; otherwise, the kernel matrix cannot capture the information between different color spaces.

\begin{figure}[!htbp]
\graphicspath{{fig/}}
\begin{center}
\includegraphics[scale=0.5]{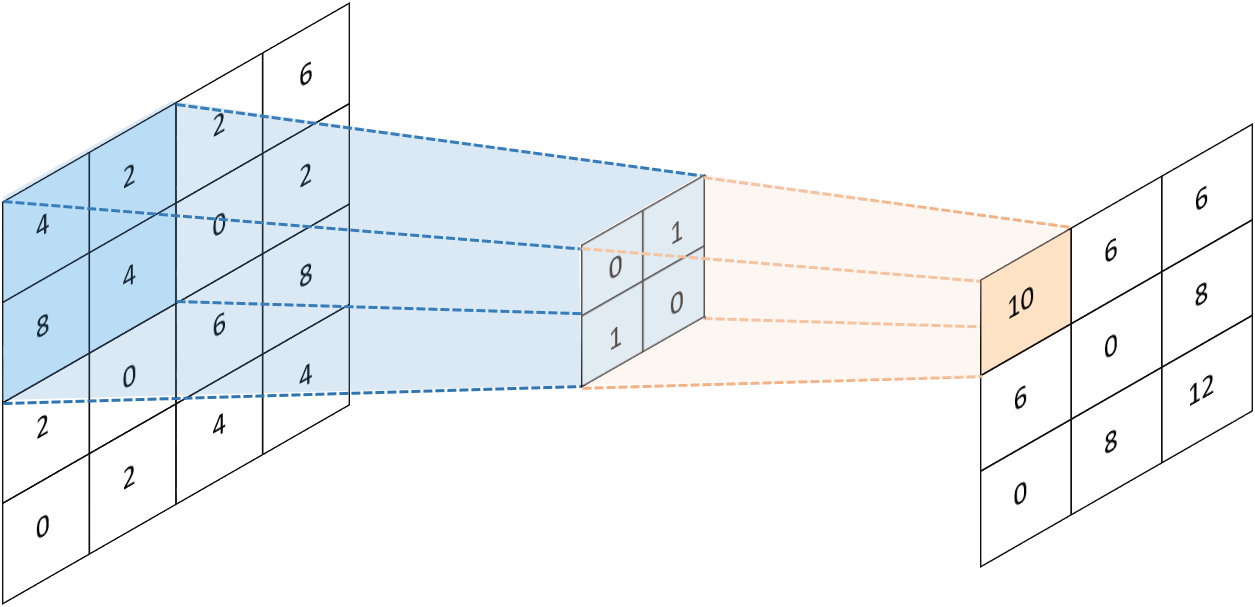}
\caption{The convolution operation.}
\end{center}
\end{figure}

Another important component is the pooling layer, also called the sub-sampling layer, which is mainly in charge of simpler tasks. The pooling layer will only retain part of the data after the convolution layer, which reduces the number of large features extracted by the convolution layer and makes the retained features more refined.

\begin{figure}[!htbp]
\graphicspath{{fig/}}
\begin{center}
\includegraphics[scale=0.5]{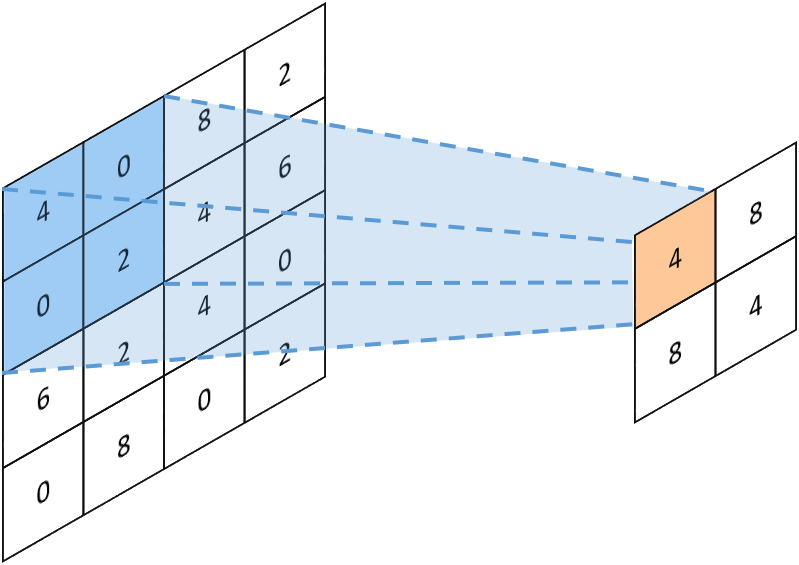}
\caption{The pooling operation.}
\end{center}
\end{figure}

Only with these two components can the convolution model be used to imitate human vision. In practical applications, the CNN model usually combines the convolution layer and pooling layer together. This is because the convolution layer often extracts a great number of features, and most of the features may be noise, which could lead to model learning in the wrong direction. This is the so-called model over-fitting problem.

Furthermore, the fully connected layers are usually connected at the end of the sequence. The function of the fully connected layer is to organize the extracted features, which were processed by the convolution and pooling layer. The correlation between the extracted features is learned in this layer.

Although the pooling layer can reduce the occurrence of over-fitting after convolution, it is inappropriate to use after the fully connected layer. Another widely known regularization technique called drop-out is designed to solve this issue. The drop-out technique randomly drops some neurons with a specific probability, and the dropped neurons are not involved in forward and backward computation. This idea directly limits the model's learning; the model can only update its parameters subject to the remaining neurons in each epoch.

\begin{figure}[!htbp]
\graphicspath{{fig/}}
\begin{center}
\includegraphics[scale=0.5]{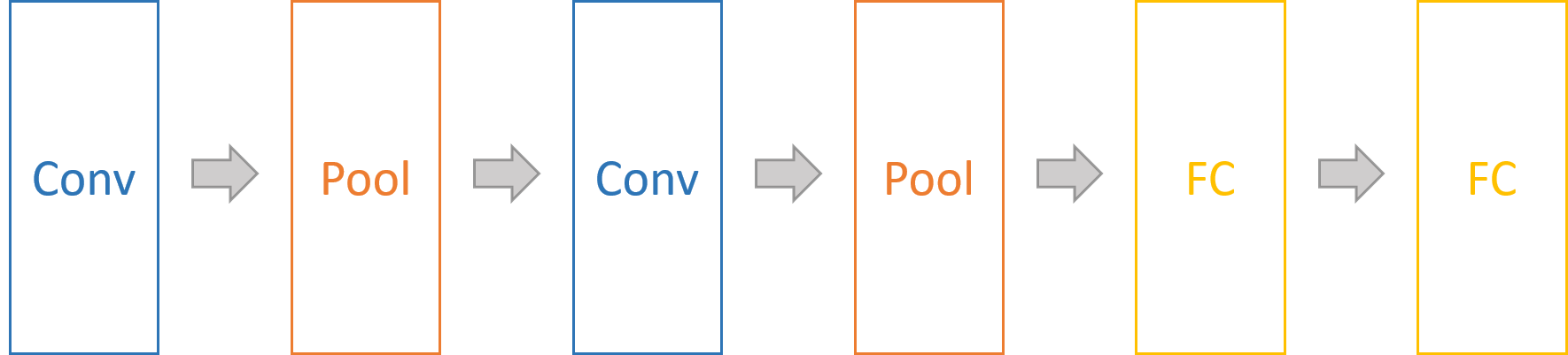}
\caption{The typical architecture of the convolution neural network, which is also the classic LeNet model.}
\end{center}
\end{figure}

Next, we introduce how to generate the data, and how to design the architecture in the first workflow.

\begin{figure}[!htbp]
\graphicspath{{fig/}}
\begin{center}
\includegraphics[scale=0.7]{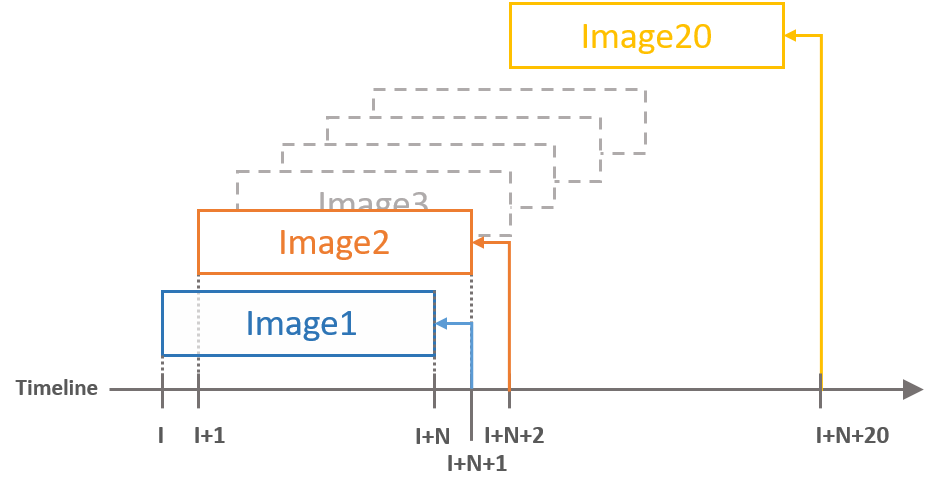}
\caption{The process of creating and labelling data in the workflow 1.}
\label{work1_1}
\end{center}
\end{figure}

The input data that we provide the computer with is a pixel image drawing from time $I$ to $I + N$, where index $I$ represents the beginning of each image and index $N$ represents the total length of the historical data we want the computer to see. After the first image is generated, the beginning of the time sequence will advance and keep generating the new images until a specific number of images has been created, meaning that the time will move from $I$ and $I + N$ to time $I + 1$ and $I + N + 1$ and proceed as thus until $M$ images have been collected. Then, because we assume increasing and decreasing patterns exist in the foreign exchange, we label the images through time $I + N + 1$, which is out of the time region of each generated image. Figure~\ref{work1_1} depicts the process of generating and labelling data in detail.

After the data are collected, we supervise the model as it learns how to classify the images into three categories: buy, sell, and not taking any action. We expect the model to predict what kind of images will rise or fall in the future; in other words, learning the data from time $I$ to $I + N$, and predicting the outcomes at time $I + N + 1$. Different from the typical image recognition problems with the CNN model, applications in finance need to make some modifications. Financial data have time-series characteristics, which cannot be captured by the convolution model. For this reason, our first workflow combines the concept of moving windows into the CNN model.

To consider the time-series properties of the financial data, the single CNN model needs to be modified. It is intuitive to think of training the new CNN model in different time regions; in more detail, we use day $I$ to $I + N + 20$ to generate data and train a convolution model. After the first run, we move to the next time window and train a new convolution model. This process continues to run until all the predictions have been made. There are two main advantages of this process: the different CNN models can capture different features in the particular time interval, and this also prevent the CNN models from using noisy features from a long time ago.

\begin{figure}[!htbp]
\graphicspath{{fig/}}
\begin{center}
\includegraphics[scale=0.7]{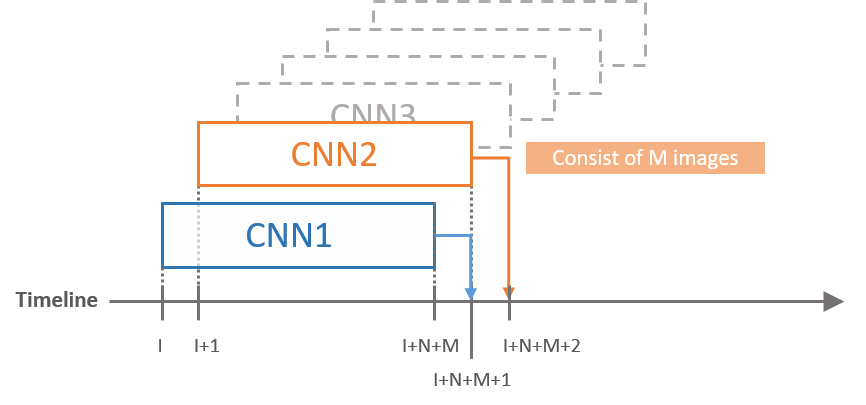}
\caption{The process that combines the moving windows into convolution model in the workflow 1.}
\end{center}
\end{figure}

For example, we may use day $1$ to day $20$ to make the data and labels, and then train a CNN model to predict the outcome on day $21$. In the second run, we use day $2$ to day $21$ to generate the new images and labels, and train a new convolution model again to predict the outcome on day $22$, and so forth.

In terms of the architecture of the convolution model, we first intend to try some simpler models, which only consist of two or three pairs of convolution and pooling layers before using the famous AlexNet model. This is because the images we want the computer to learn are simple sets of one to four closed price line plots including high, low, and the moving average. They are not as complex as the ImageNet Challenge.

\begin{figure}[!htbp]
\graphicspath{{fig/}}
\begin{center}
\includegraphics[scale=0.7]{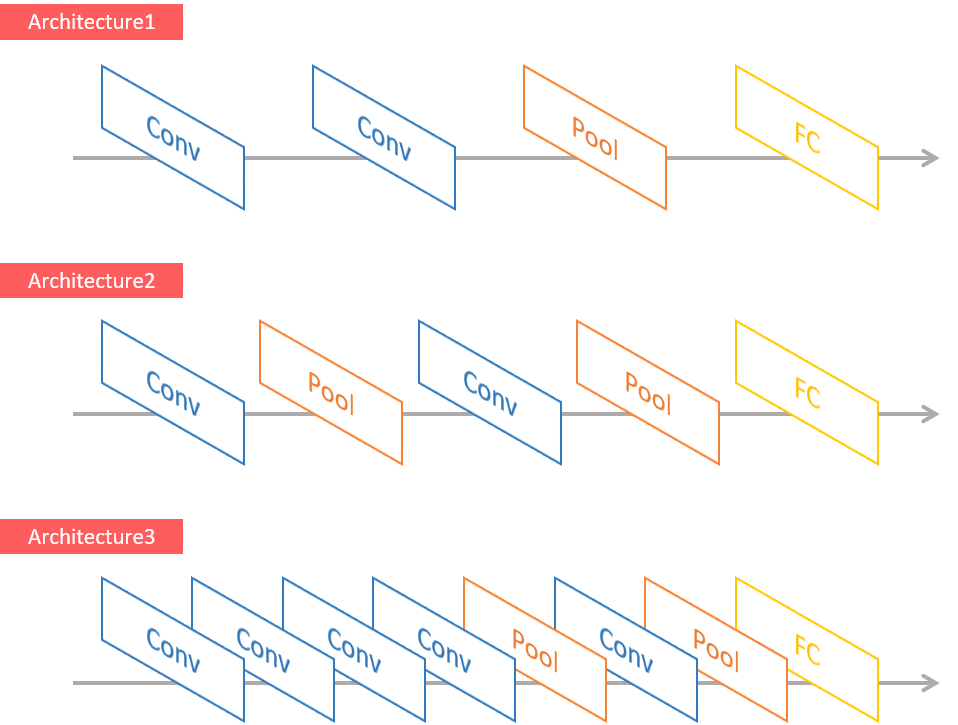}
\caption{The three architectures in the workflow 1.}
\label{arch3}
\end{center}
\end{figure}

All the architectures we used are shown in Figure~\ref{arch3}, where Conv, Pool, and FC are the convolutional layer, pooling layer, and fully connected layer, respectively.

\begin{enumerate}
\item Architecture 1: Conv + Conv + Pool + FC
\item Architecture 2: (Conv + Pool) * 2 + FC
\item Architecture 3: Conv * 4 + Pool + Conv + Pool + FC
\end{enumerate}

In the first architecture, we used two convolution layers, further connected to a pooling layer and a fully connected layer. In the second architecture, we used two pairs of convolution and pooling layers and a fully connected layer, which is similar to the architecture of LeNet. We expect that these two simple architectures can enable the computer to learn the simple structure from the input images.

In the third architecture, we designed a deeper architecture consisting of more convolution layers. We used this architecture because we tried to solve the under-fitting problem from the model; simple architecture was not sufficient to learn features from input images.

The results of these experiments do not fit the expectation; whether simple or complex, the architectures do not fit the convolution model well. The experimental procedures are illustrated in detail in the Experimental Results section.

Another architecture which is widely used in our second workflow is the well-known AlexNet model. The AlexNet model appeared in 2012, beat the previous champion, and became the state-of-the-art model in the ImageNet Challenge, which has over 15M labelled high resolution images and roughly 22,000 categories. The AlexNet model has a deeper structure than LeNet, containing five convolutional layers, three fully connected layers, and a softmax layer. To prevent the model from over-fitting, the AlexNet model also uses a new regularization method called drop-out and data augmentation, which horizontally flips the image or performs random cropping. The AlexNet model also uses the ReLU as the activation function, which is a non-saturating activation function and is several times faster than tanh and sigmoid units. With these improvements and excellent GPU support, the AlexNet model has become one of the most powerful models today.

\section{Methodology}
In this section, we introduce the architectures we used in our experiments and justify our decision for using these workflows. We also illustrate some data preprocessing techniques used to generate our inputs. The deep learning frameworks used in each workflow are the Python Keras module and NVIDIA Digits with the Caffe back-end. All the convolution models in both workflows were trained for $30$ epochs and were speeded up by the GTX TITAN GPU. We also tried to observe the result of different epochs, even up to $4000$ epochs, but the over-fitting almost significantly occurs at about $50–--100$ epochs. The workflows are as follows:

\subsection{Workflow 1}
\begin{figure}[!htbp]
\graphicspath{{fig/}}
\begin{center}
\includegraphics[scale=0.7]{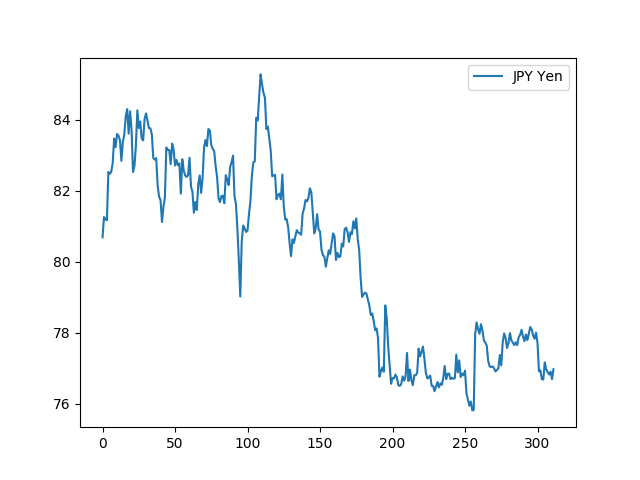}
\caption{The exchange rates of Japanese Yen from November 9, 2010 to January 13, 2011.}
\label{fig1}
\end{center}
\end{figure}

In the first workflow, we used the real-world exchange rates of Japanese Yen from 2010 to 2011. We designed three kinds of convolution architectures and expected one of these architectures to fit the real-world data well. The overview of the raw data is shown in Figure~\ref{fig1} and the first workflow is enumerated in detail as follows:

\begin{enumerate}
\item Transform the quantitative price data to image data using the Python Matplotlib module, and create classification-based labels which consist of buy, sell, and not taking any action.
\item Create the three architectures of the CNN model by using the Python Keras deep learning module. Each of the architectures will be experimented independently.
\item Train the CNN model and tweak the parameters to maximize accuracy. The number of epochs used for training ranges from $30$ to $100$.
\item Evaluate the model with a confusion matrix for currency performance.
\item Repeat the above steps until the best model is found.
\end{enumerate}

\subsection{Workflow 2}
\begin{figure}[!htbp]
\graphicspath{{fig/}}
\begin{center}
\includegraphics[scale=0.5]{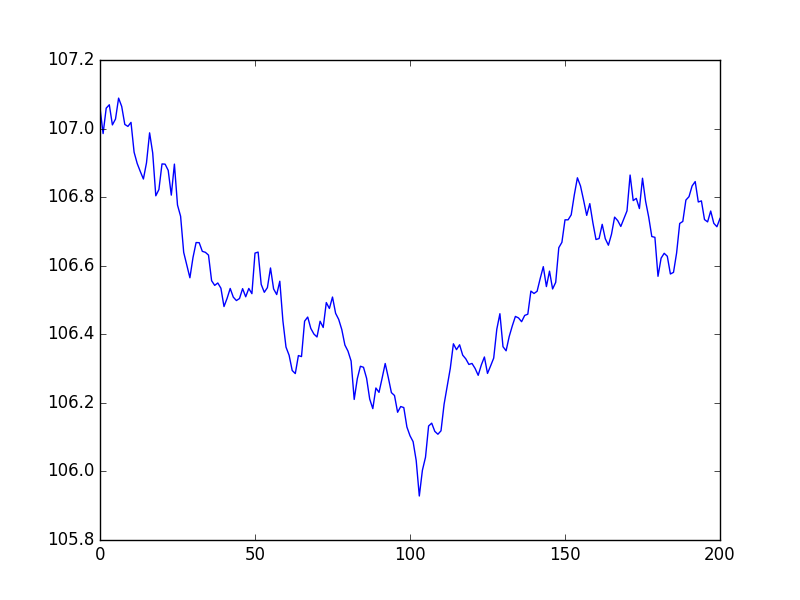}
\caption{One of the simulation data generated by the Geometric Brownian Motion process (GBM).}\label{simulation}
\end{center}
\end{figure}

Because the performance of workflow 1 was not as good as expected, we switched to using simulation data from the GBM. We simulated $90$ days foreign exchange rate data repeatedly, for $100$ times, with a $1\%$ yearly return and $25\%$ yearly standard errors. We believed these prices approximately followed a subset of the real-world data, and therefore, we expected the new architecture to fit well in the subset of the real world. One of the simulated data is shown in Figure~\ref{simulation} and the second workflow is enumerated in detail as follows:
\begin{enumerate}
\item Transform the quantitative price data to image data using the Python Matplotlib module, and create classification-based labels which consist of buy, sell, and not taking any action.
\item Create the AlexNet architecture of the CNN model by using NVIDIA DIGITS with the Caffe back-end. NVIDIA DIGITS is a lightweight tool, especially good at presenting the training process in real time.
\item Train the AlexNet model and tweak the parameters to maximize accuracy. The number of epochs used for training is $50$.
\item Evaluate the model with a confusion matrix for currency performance.
\item Repeat the above steps until the best model is found.
\end{enumerate}

The structure of workflow 2 is almost the same as that of workflow 1. The main difference in workflow 2 is the way data is labelled; in workflow 1, the same strategy is used to label all inputs, but many kinds of strategies are used in workflow 2. In workflow 2, we used the AlexNet model with its default parameters, and tweaked only the epochs and the different kinds of input images. The strategies we used in the workflow 2 are listed below:
\begin{enumerate}
\item Use every $20$ days period as an image and the following $5$ days as holding days; that is, we use day $1$ to day $20$ as the input image, and use day $25$ to label day $20$. If the price on day $25$ is greater than day $20$ by at least $1\%$, then we will buy on day $20$ and sell on day $25$. If the price on day $25$ is less than day $20$ by at least $1\%$, then we will sell on day $20$ and buy on day $25$. Otherwise, no action will be taken.
\item In this case, we tried to use the moving average as our strategy. Because we wanted the inputs to be more distinguishable by the model, the rule we used was that if MA5 is greater than MA7 by at least $1\%$ and MA7 is greater than MA10 by at least $1\%$ on day $9$, then we will buy on day $6$ and sell on day $9$. If MA5 is less than MA7 by at least $1\%$ and MA7 is less than MA10 by at least $1\%$ on day $9$, then we will sell on day $6$ and buy on day $9$. Otherwise, no action will be taken.
\item Furthermore, we also simulated both open and closed price, and plotted it with the MA5, MA10, and MA20 lines. We used every $15$ days period as the image and the following $5$ days as the holding period. The strategy used here is that if the opening price on day $20$ is greater than the closing price on day $15$ by at least $2\%$, then we will buy on day $15$ and sell on day $20$. If the opening price on day $20$ is less than the closing price on day $15$ by at least $1\%$, then we will sell on day $15$ and buy on day $20$.
\end{enumerate}

\section{Experimental Results}
First, we introduce three ways to pre-process the image data; second, we discuss problems we encountered in the experimental procedure and illustrate how to solve them. The pre-process frameworks we used are the Python Matplotlib module and Python Pillow module.
The following are the three ways in which we pre-process our images:

\begin{enumerate}
\item Crop the images without the information of the x-axis and y-axis. This is because we want our input data to be as clean as possible.
\item Use the RGB color space to capture the information of moving average lines. Different colors will be given to each moving average line, so the moving average lines will be represented in the different channels.
\item Invert the color space to highlight only the lines in the image. The background will become black, which means the value of each background pixel is zero.
\end{enumerate}

\begin{figure}[!htbp]
\graphicspath{{fig/}}
\begin{center}
\includegraphics[scale=0.9]{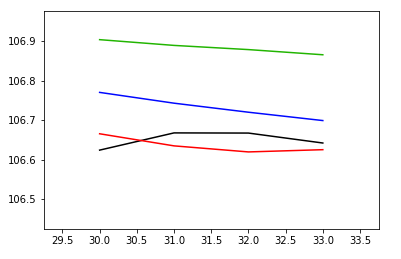}
\caption{The image data with price and moving average lines without resizing. Black is the price line, red is the MA5 line, blue is the MA10 line, and green is the MA20 line. There are still many different permutations and combinations of the price and moving average lines.}
\label{fig_d1}
\end{center}
\end{figure}

\begin{figure}[!htbp]
\graphicspath{{fig/}}
\begin{center}
\includegraphics[scale=0.9]{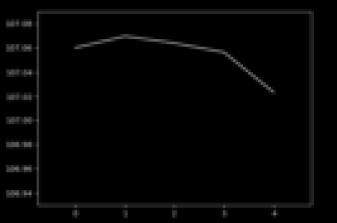}
\caption{The image data only with price line preprocessed with inversion.}
\label{fig_d2}
\end{center}
\end{figure}

The moving average lines we used are MA5, MA7, MA10, and MA20. We used moving average lines to simulate our inputs and increase similarity to the trading charts. We also rescaled the images to different sizes, for example, $100\times 150$ or $300\times 400$. We also tried to set the different y-axes in the same scale. The image of the moving average lines is shown in Figure~\ref{fig_d1} and the inverted one is shown in Figure~\ref{fig_d2}.


\subsection{Workflow 1}
In workflow 1, we tried three architectures. The default time region is 20; in each region, we used every 5-day period to create the image data and used the next day to label the input images. Each architecture used the framework of the moving windows and predicted a hundred times. The three architectures we tried are as follows:

\begin{enumerate}
\item Architecture 1: Conv + Conv + Pool + FC
\item Architecture 2: (Conv + Pool) * 2 + FC
\item Architecture 3: Conv * 4 + Pool + Conv + Pool + FC
\end{enumerate}

We used the first two architectures (architecture 1 and architecture 2) because we expected a simple model could solve our problem; however, the results were not good. Therefore, we next used a deeper structure with architecture 3; we added more convolution layers and filters in the first two layers to help the model extract more detailed information. We hoped a more complex architecture would help solve this problem. Unfortunately, neither the simple nor the complex architectures worked well. The complex one did not improve the performance of classification. The experimental results of each architecture are shown below.

\subsection{Architecture 1}\label{Architecture 1}
For architecture 1, we carried out three experiments. We inverted all input images and resized them to $100\times 150$. We used different parameters in each experiment, as follows:
\begin{enumerate}
\item In the first experiment, we used a kernel size of $30\times 40$, with $5$ kernels and $128$ fully connected units. The pooling layer we used was MaxPooling $2\times 2$ and the time region used was $20$, which means using $20$ days historical information to predict the action for the next day.
\item In the second experiment, we used a kernel size of $30\times 40$, with $10$ kernels and $128$ fully connected units. The pooling layer we used was MaxPooling $2\times 2$ and the time region used was $20$.
\item In the third experiment, we used a kernel size of $30\times 40$, with $5$ kernels and $128$ fully connected units. The pooling layer we used was MaxPooling $2\times 2$, but this time we used $30$ days as our time region, which means using $30$ days historical information to predict the action for the next day.
\end{enumerate}

\begin{figure}[!htbp]
\graphicspath{{fig/}}
\begin{center}
\includegraphics[scale=0.5]{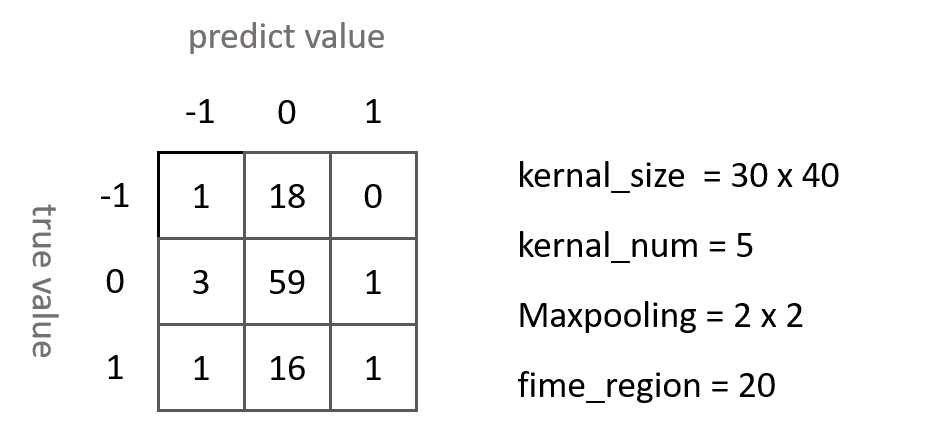}
\caption{The confusion matrix of the experiment 1 in architecture 1.}
\label{c1}
\end{center}
\end{figure}

\begin{figure}[!htbp]
\graphicspath{{fig/}}
\begin{center}
\includegraphics[scale=0.5]{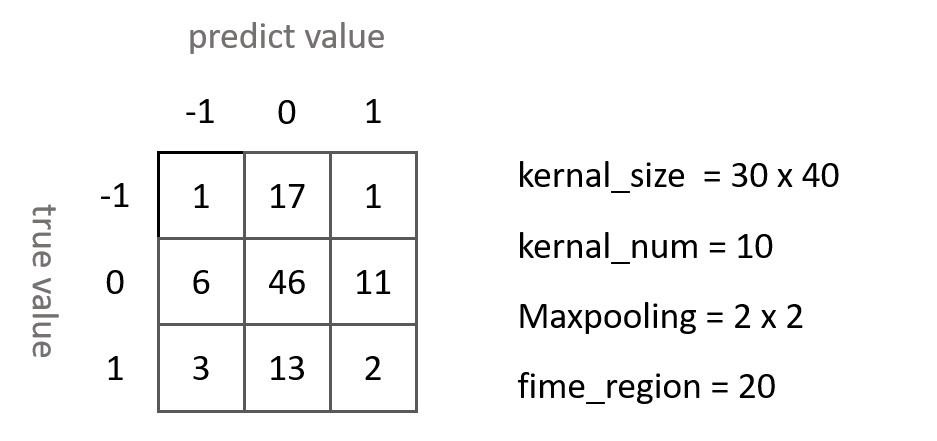}
\caption{The confusion matrix of the experiment 2 in architecture 1.}
\label{c2}
\end{center}
\end{figure}

\begin{figure}[!htbp]
\graphicspath{{fig/}}
\begin{center}
\includegraphics[scale=0.5]{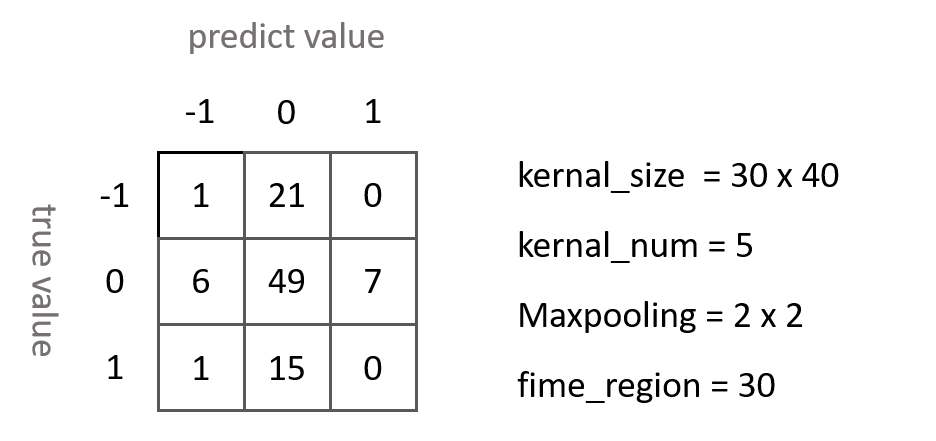}
\caption{The confusion matrix of the experiment 3 in architecture 1.}
\label{c3}
\end{center}
\end{figure}

The results of the three experiments are described in Figures~\ref{c1}---\ref{c3}, respectively. There is no significant improvement between parameters; the model often predicts the action to be doing nothing.


\subsection{Architecture 2}\label{Architecture 2}
\begin{figure}[!htbp]
\graphicspath{{fig/}}
\begin{center}
\includegraphics[scale=0.5]{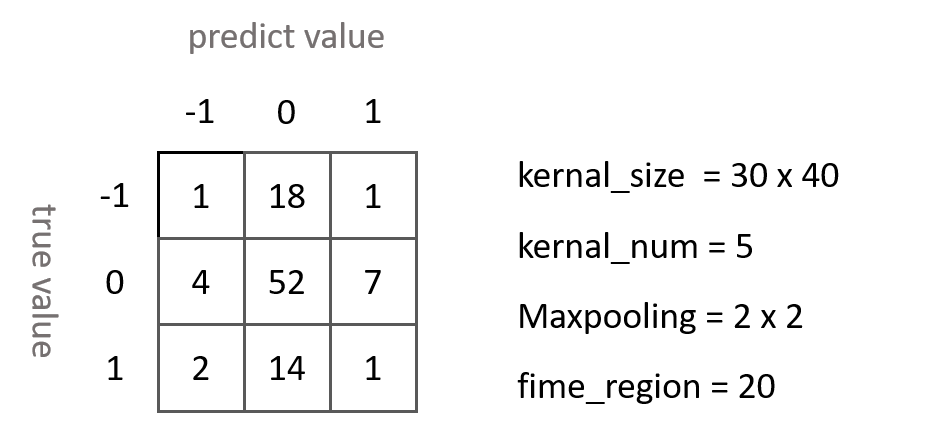}
\caption{The confusion matrix of the experiment 1 in architecture 2.}
\label{c4}
\end{center}
\end{figure}

The parameters used in the second experiment are the same as those used in the first experiment; only the architecture of the model is different. The performance of the second architecture is also poor, with the model once again giving the prediction of taking no action often. One result is shown in Figure~\ref{c4}.

We made some changes to the architectures because we obtained poor performance with the architectures and experiments above: we added two more convolution layers and an additional pooling layer to make the model deeper and more complex.

\subsection{Architecture 3}\label{Architecture 3}
\begin{figure}[!htbp]
\graphicspath{{fig/}}
\begin{center}
\includegraphics[scale=0.5]{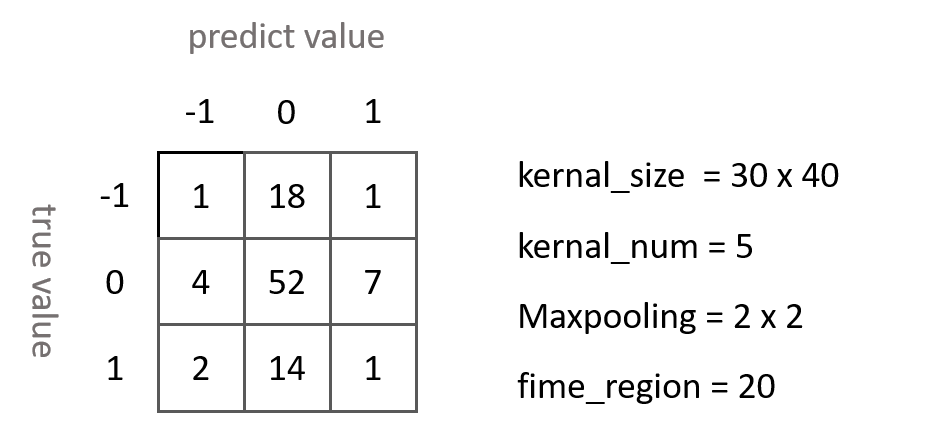}
\caption{The confusion matrix of the experiment 1 in architecture 3.}
\label{c5}
\end{center}
\end{figure}

With the new, more complex architecture, we designed three kinds of experiments. The parameters of each experiment are almost the same; the only difference between the three experiments is the number of kernels. This is because we expected more filters would capture more features of the image. In experiments 1–--3, the number of kernels is designated as $5, 10$ and $20$. The results of each experiment are described in Figures~\ref{c5}--–\ref{c7}.

\begin{figure}[!htbp]
\graphicspath{{fig/}}
\begin{center}
\includegraphics[scale=0.5]{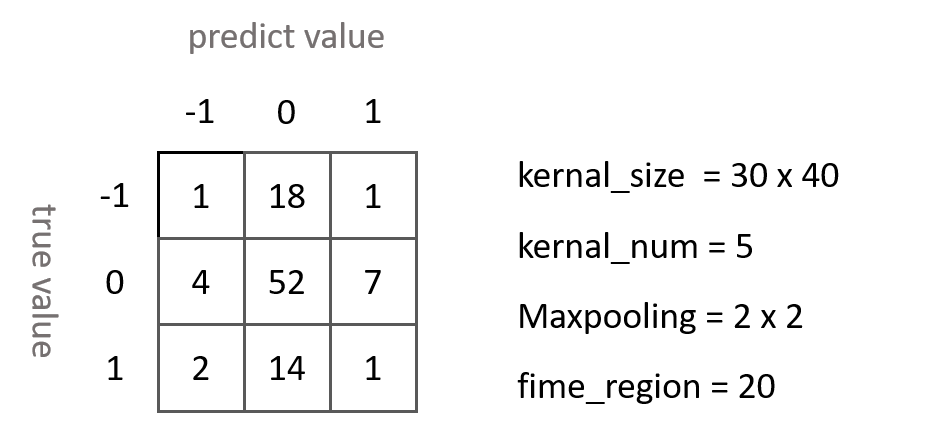}
\caption{The confusion matrix of the experiment 2 in architecture 3.}
\label{c6}
\end{center}
\end{figure}
\begin{figure}[!htbp]
\graphicspath{{fig/}}
\begin{center}
\includegraphics[scale=0.5]{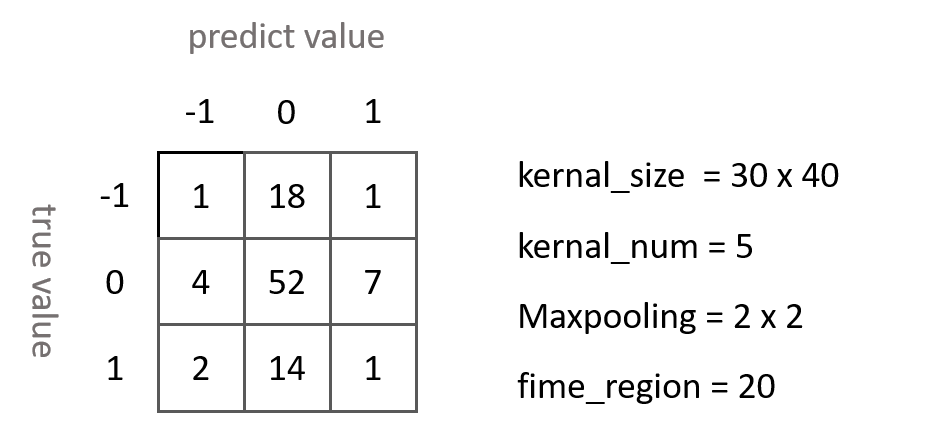}
\caption{The confusion matrix of the experiment 3 in architecture 3.}
\label{c7}
\end{center}
\end{figure}

From the results of the three architectures, we can clearly see that none of the experiments yielded good performance. Additionally, each model is unstable due to over-fitting. This is because the number of input images is too small to train the convolution model; if the time region is $20$ and if we use every $5$ days period to create an image, we only have $16$ images of training data.

The convolution model can fit the given $16$ images training data well, but cannot recognize images with many differences to the training data.
The only way to obtain more real-world training data is to extend the time region; in finance, however, older information does not help predict future data. Additional data would only increase the occurrence of noise, meaning we cannot simply extend the time region to collect more training data; an alternative approach is required.

\subsection{Workflow 2}
Before addressing the real-world data, we wanted to fit the model with the simulated data. This is because the simulated data can give us sufficient data, with little noise. In addition, simulated data accurately represents a subset of the real-world data, and therefore may be easier to fit. If we can fit the small world well, the convolution model can learn strategies from it. We used a mean of $1\%$ and a standard error of $25\%$ to simulate $90$ days data; we simulated it many times to generate enough data for the convolution model. The three experiments, trained with the simulated data, are introduced in detail as follows:

\subsection{Experiments 1}\label{experiment 1}
\begin{figure}[!htbp]
\graphicspath{{fig/}}
\begin{center}
\includegraphics[scale=0.5]{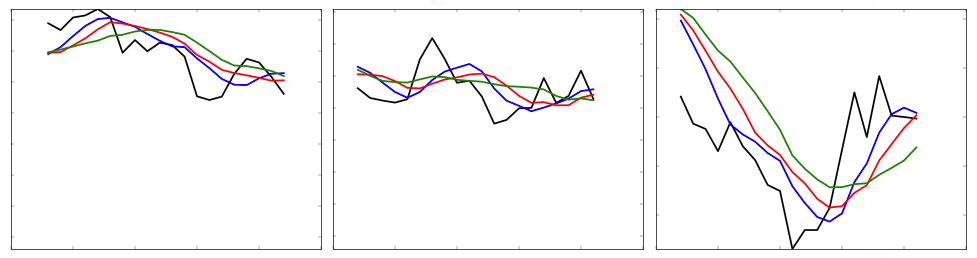}
\caption{The experiment 1 with label $1$.}
\label{label1_wf2_1}
\end{center}
\end{figure}

\begin{figure}[!htbp]
\graphicspath{{fig/}}
\begin{center}
\includegraphics[scale=0.5]{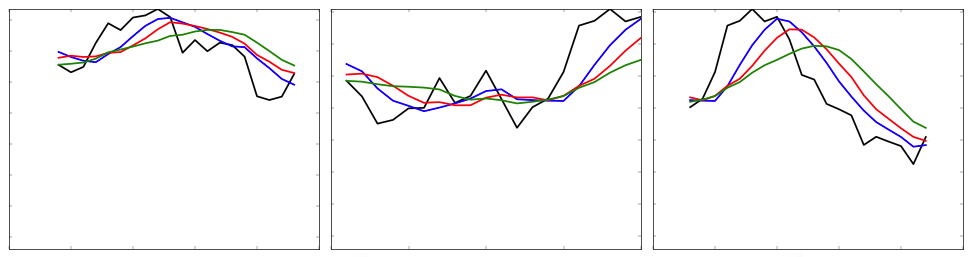}
\caption{The experiment 1 with label $-1$.}
\label{label-1_wf2_1}
\end{center}
\end{figure}

\begin{figure}[!htbp]
\graphicspath{{fig/}}
\begin{center}
\includegraphics[scale=0.5]{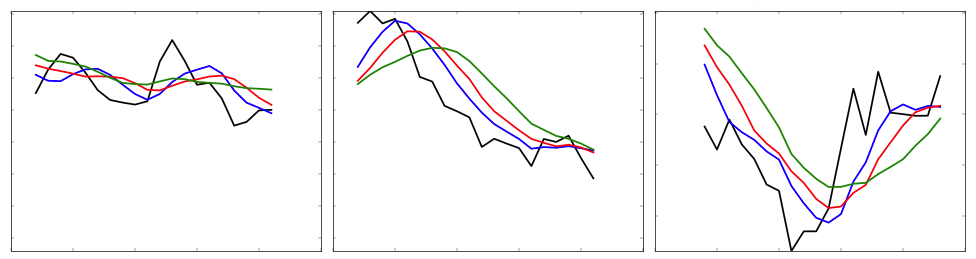}
\caption{The experiment 1 with label $0$.}
\label{label0_wf2_1}
\end{center}
\end{figure}

In experiment 1, we used every $20$ days period to create an image and the following $5$ days as the holding days; that is, we may use day $1$ to day $20$ as the input image, and day $25$ to label day $20$. If the price on day $25$ is larger than day $20$ by at least $1\%$, then we will buy on day $20$ and sell on day $25$. If the price on the day $25$ is smaller than day $20$ by at least $1\%$, then we will sell on day $20$ and buy on day $25$. Otherwise, no action will be taken. The images of the three different classes are shown in Figures~\ref{label1_wf2_1}---\ref{label0_wf2_1}.


\begin{figure}[!htbp]
\graphicspath{{fig/}}
\begin{center}
\includegraphics[scale=0.9]{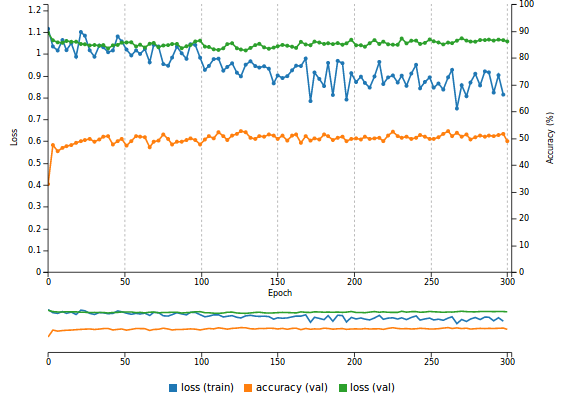}
\caption{The training process of the experiment 1.}
\label{trainprocess_w2_1}
\end{center}
\end{figure}

We can clearly see that each class cannot be easily distinguished by humans; this also makes it difficult for the convolution model to recognize the pattern of each class. In the training process of this case, which is shown in Figure~\ref{trainprocess_w2_1}, the loss of the training data and the validation data were not decreasing. The over-fitting problem also occurred after the $100$th epoch.

\begin{figure}[!htbp]
\graphicspath{{fig/}}
\begin{center}
\includegraphics[scale=0.9]{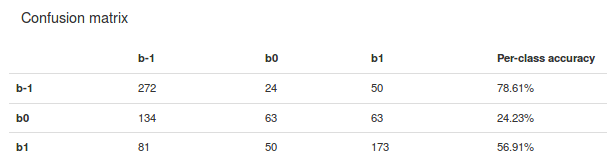}
\caption{The training process of the experiment 1.}
\label{train_confusion_w2_1}
\end{center}
\end{figure}

\begin{figure}[!htbp]
\graphicspath{{fig/}}
\begin{center}
\includegraphics[scale=0.9]{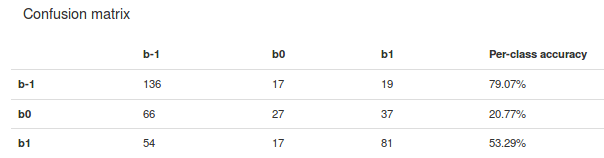}
\caption{The training process of the experiment 1.}
\label{test_confusion_w2_1}
\end{center}
\end{figure}

This time, the accuracy of the simple convolution model is better than the moving average one. The model predicts better in label $1$ and $-1$, but there are still many regions in which it could be improved. Figures~\ref{train_confusion_w2_1} and~\ref{test_confusion_w2_1} show the confusion matrix of the training and testing data.


\subsection{Experiments 2}\label{experiment 2}
Inspired by experiment 1, we tried to use the moving average as our strategy. Because we wanted the inputs to be more distinguishable by the model, the rule we used was that if MA5 is greater than MA7 by at least $1\%$ and MA7 is greater than MA10 by at least $1\%$ on day $9$, then we will buy on day $6$ and sell on day $9$. If MA5 is less than MA7 by at least $1\%$ and MA7 is less than MA10 by at least $1\%$ on day $9$, then we will sell on day $6$ and buy on day $9$. Otherwise, no action will be taken. 

\begin{figure}[!htbp]
\graphicspath{{fig/}}
\begin{center}
\includegraphics[scale=0.5]{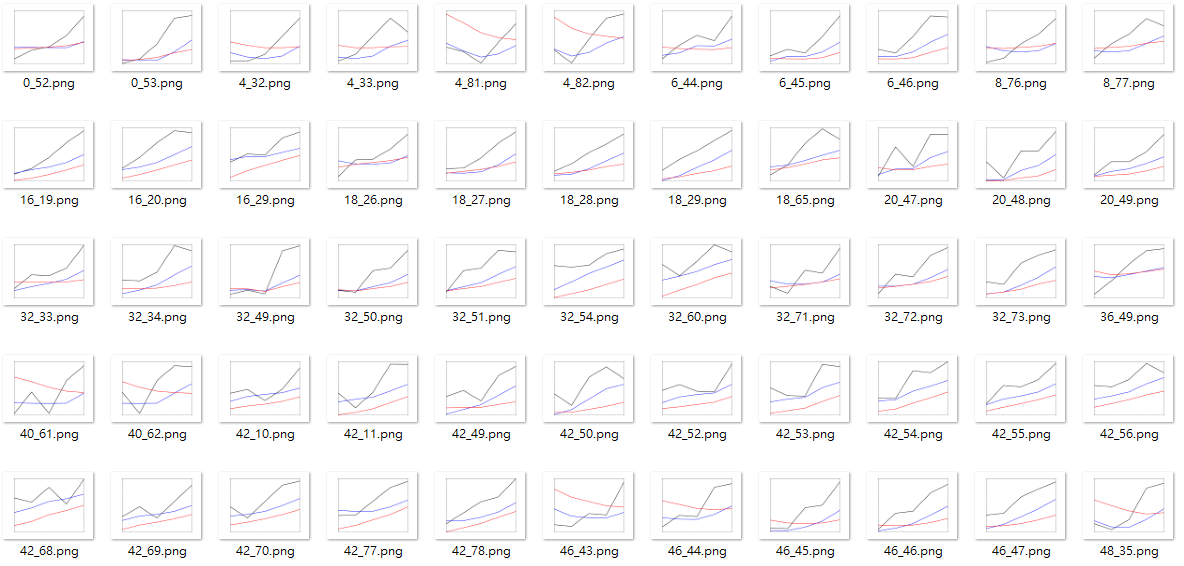}
\caption{The experiment 2 with label $1$.}
\label{label1}
\end{center}
\end{figure}

The three kinds of labelled images are shown in Figures~\ref{label1}--–\ref{label0}, and we can see that the pattern is more significant in the buying $(1)$ and selling $(-1)$ labels now. This makes it easier for the convolution model to detect the difference between the strategies. After the trials of experiment 3, we achieved an accuracy rate of $82\%$, which is a significant improvement over experiments 1 and 2. We also scaled the images to the maximum and minimum of the prices and the moving average; this yielded an $80\%$ accuracy rate.

\begin{figure}[!htbp]
\graphicspath{{fig/}}
\begin{center}
\includegraphics[scale=0.5]{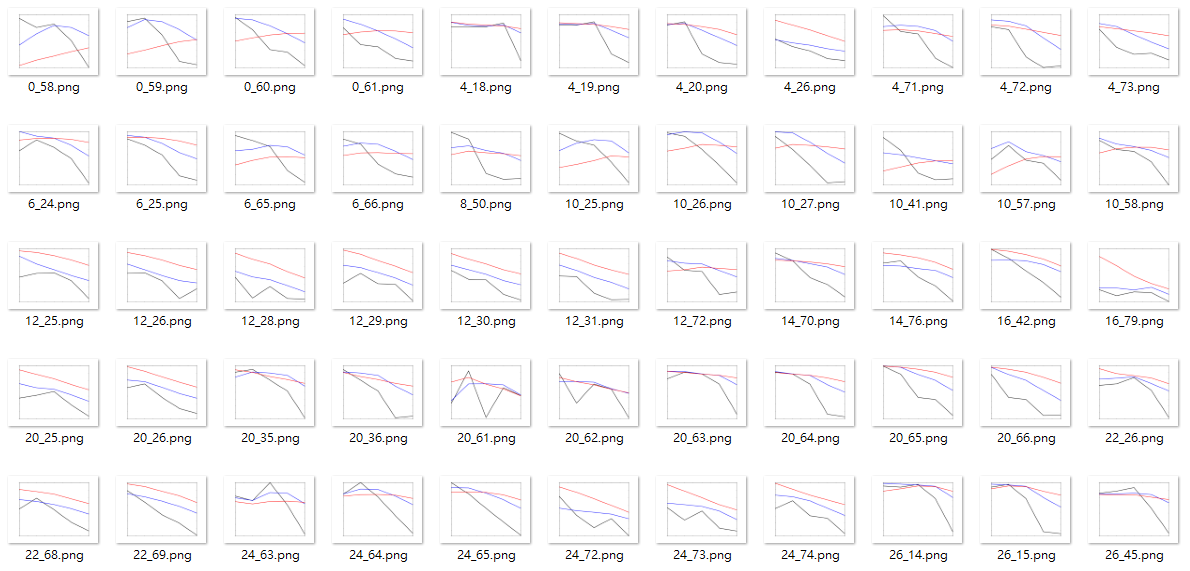}
\caption{The experiment 2 with label $-1$.}
\label{label-1}
\end{center}
\end{figure}
\begin{figure}[!htbp]
\graphicspath{{fig/}}
\begin{center}
\includegraphics[scale=0.5]{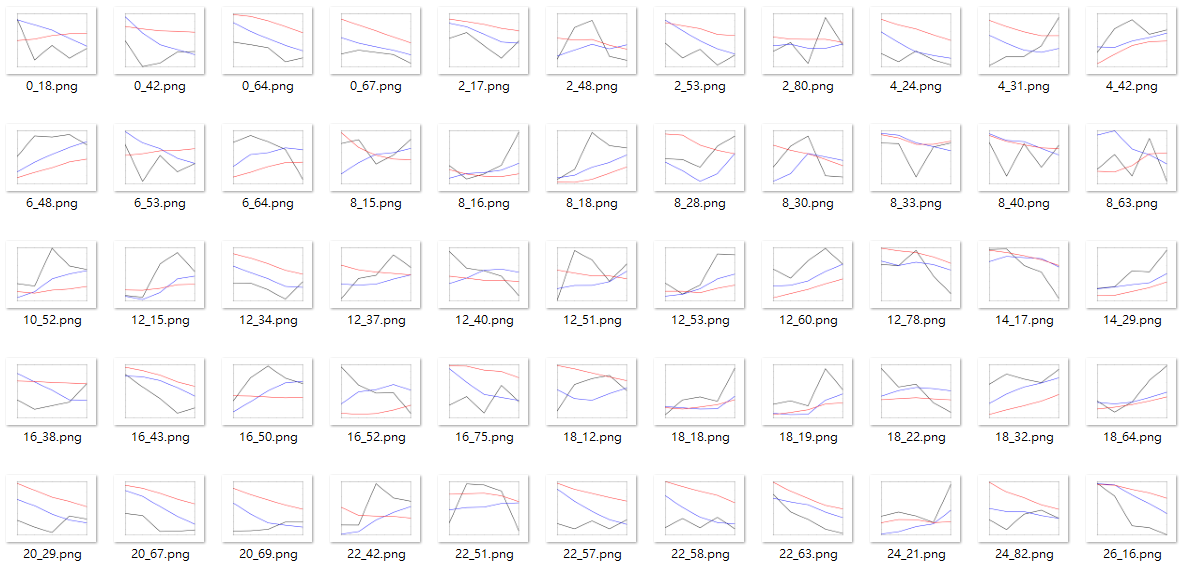}
\caption{The experiment 2 with label $0$.}
\label{label0}
\end{center}
\end{figure}

\begin{figure}[!htbp]
\graphicspath{{fig/}}
\begin{center}
\includegraphics[scale=0.7]{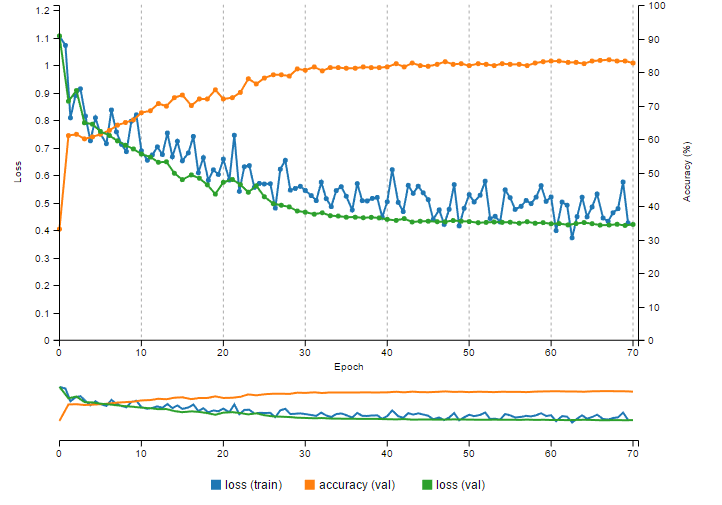}
\caption{The training process of the experiment 3.}
\label{train_process}
\end{center}
\end{figure}

The training process is shown in Figure~\ref{train_process}. In the experiment, we used $25\%$ of the data for validation and $25\%$ for testing. The accuracy rate increased to $82\%$ in the $70$th epoch. The problem of over-fitting does not occur, which can be explained by the loss of the training data and the validation data.

\begin{figure}[!htbp]
\graphicspath{{fig/}}
\begin{center}
\includegraphics[scale=0.7]{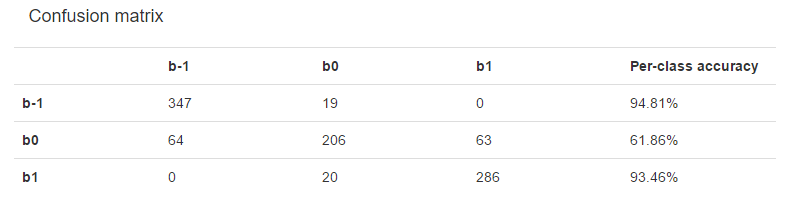}
\caption{The confusion matrix of the training data.}
\label{train_confusion}
\end{center}
\end{figure}

\begin{figure}[!htbp]
\graphicspath{{fig/}}
\begin{center}
\includegraphics[scale=0.7]{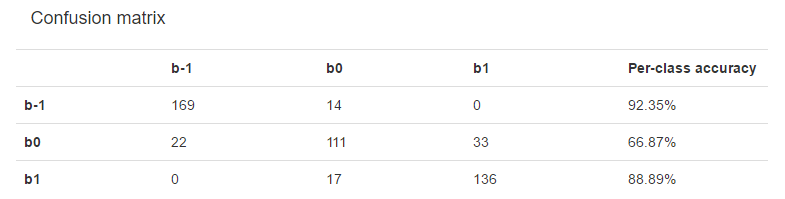}
\caption{The confusion matrix of the testing data.}
\label{test_confusion}
\end{center}
\end{figure}

The confusion matrix of the training data and the testing data are shown in Figures~\ref{train_confusion} and~\ref{test_confusion}. From the result, we can see that the accuracy of each class is not significantly impacted by the over-fitting problem. The accuracy of the testing data is only slightly lower than the training data.


\begin{figure}[!htbp]
\graphicspath{{fig/}}
\begin{center}
\includegraphics[scale=0.7]{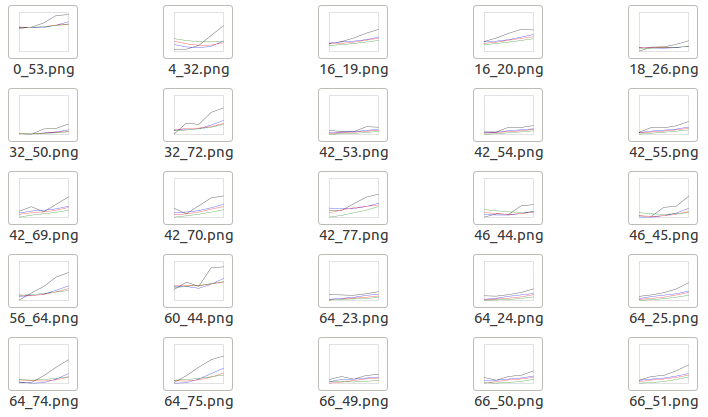}
\caption{The experiment 2 with scaling with label $1$.}
\label{label1_wy}
\end{center}
\end{figure}

The experimental results of the images scaled to the maximum and minimum of the prices and moving average are as follows. Figures~\ref{label1_wy}--–\ref{label0_wy} show the images classified by the MA strategy. Figure~\ref{train_process_wy} describes the training process. Figures~\ref{train_confusion_wy} and~\ref{test_confusion_wy} show the confusion matrix of the training and the testing data. The results of this case achieved an accuracy rate of $82\%$, which is better compared to the earlier rate.

\begin{figure}[!htbp]
\graphicspath{{fig/}}
\begin{center}
\includegraphics[scale=0.7]{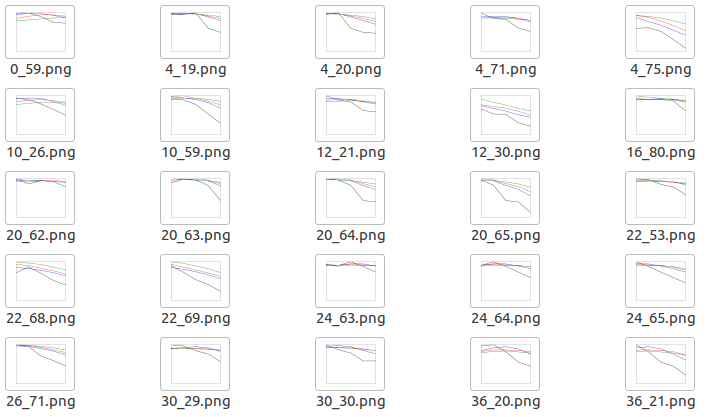}
\caption{The experiment 2 with scaling with label $-1$.}
\label{label-1_wy}
\end{center}
\end{figure}
\begin{figure}[!htbp]
\graphicspath{{fig/}}
\begin{center}
\includegraphics[scale=0.7]{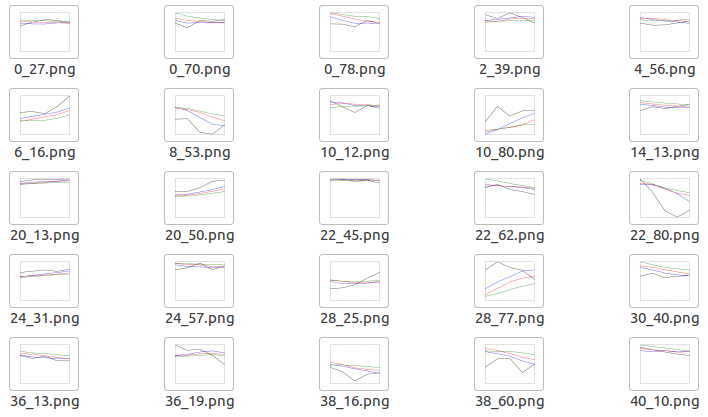}
\caption{The experiment 2 with scaling with label $0$.}
\label{label0_wy}
\end{center}
\end{figure}
\begin{figure}[!htbp]
\graphicspath{{fig/}}
\begin{center}
\includegraphics[scale=0.7]{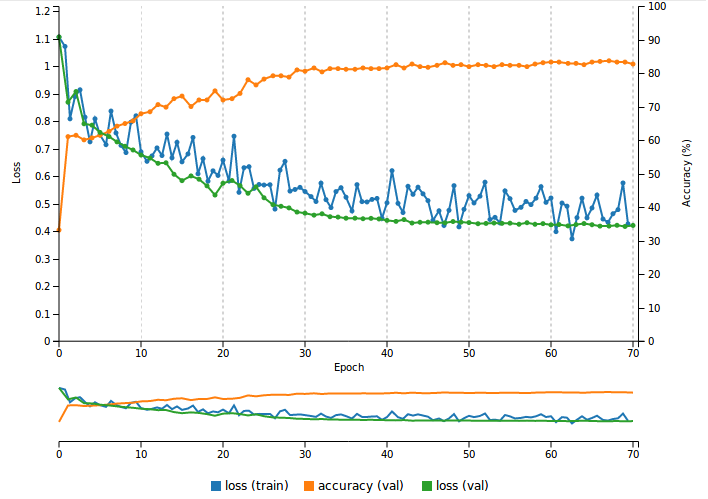}
\caption{The training process of the experiment 2 with scaling.}
\label{train_process_wy}
\end{center}
\end{figure}
\begin{figure}[!htbp]
\graphicspath{{fig/}}
\begin{center}
\includegraphics[scale=0.7]{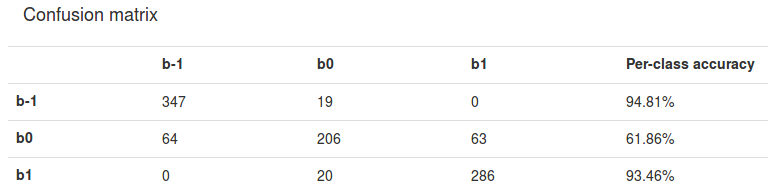}
\caption{The confusion matrix of the training data with scaling.}
\label{train_confusion_wy}
\end{center}
\end{figure}
\begin{figure}[!htbp]
\graphicspath{{fig/}}
\begin{center}
\includegraphics[scale=0.5]{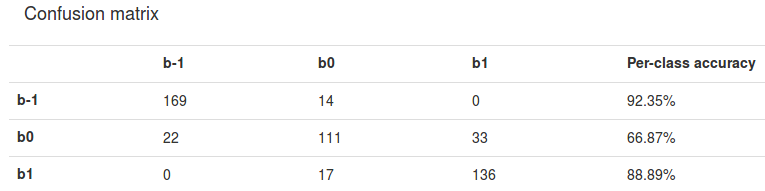}
\caption{The confusion matrix of the testing data with scaling.}
\label{test_confusion_wy}
\end{center}
\end{figure}

\subsection{Experiments 3}\label{experiment 3}
\begin{figure}[!htbp]
\graphicspath{{fig/}}
\begin{center}
\includegraphics[scale=0.7]{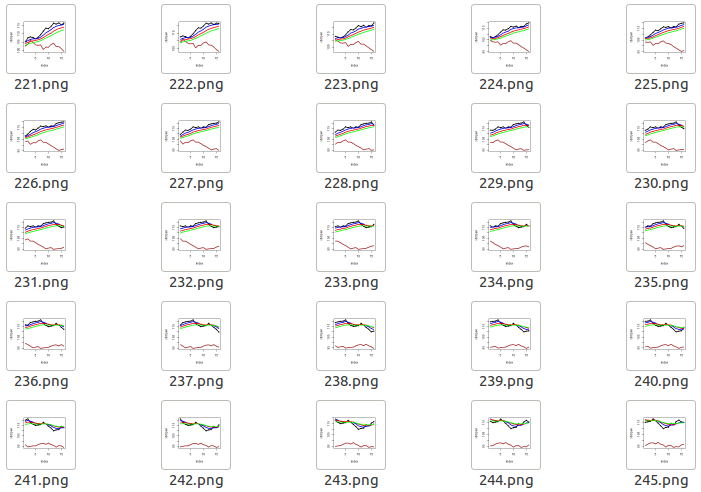}
\caption{The experiment 3 using open, close, MA5, MA10, and MA20 with label $1$.}
\label{label1_pecu}
\end{center}
\end{figure}

We also simulated both open and closed price, and plotted them with the MA5, MA10, and MA20 lines. We used every $15$ days period to create an image and the following $5$ days as the holding days. If the opening price on day $20$ is greater than the closing price on day $15$ by at least $2\%$, then we will buy on day $15$ and sell on day $20$. If the opening price on day $20$ is less than the closing price on day $15$ by at least $1\%$, then we will sell on day $15$ and buy on day $20$. The three kinds of labelled images are shown in Figures~\ref{label1_pecu}--–\ref{label0_pecu}.

\begin{figure}[!htbp]
\graphicspath{{fig/}}
\begin{center}
\includegraphics[scale=0.7]{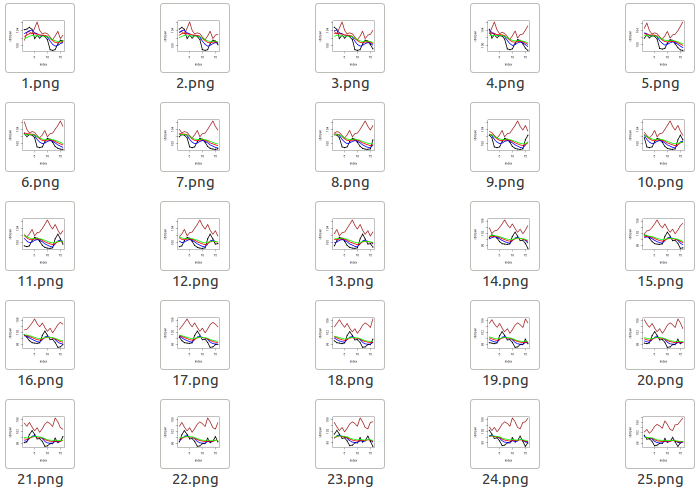}
\caption{The experiment 3 using open, close, MA5, MA10, and MA20 with label $-1$.}
\label{label-1_pecu}
\end{center}
\end{figure}
\begin{figure}[!htbp]
\graphicspath{{fig/}}
\begin{center}
\includegraphics[scale=0.7]{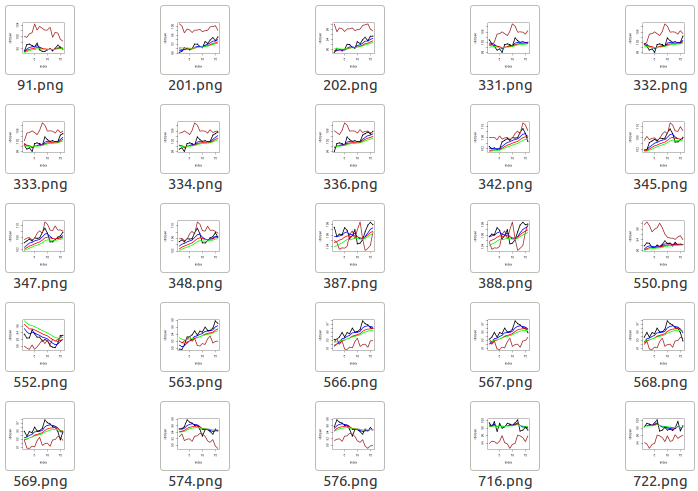}
\caption{The experiment 3 using open, close, MA5, MA10, and MA20 with label $0$.}
\label{label0_pecu}
\end{center}
\end{figure}

\begin{figure}[!htbp]
\graphicspath{{fig/}}
\begin{center}
\includegraphics[scale=0.8]{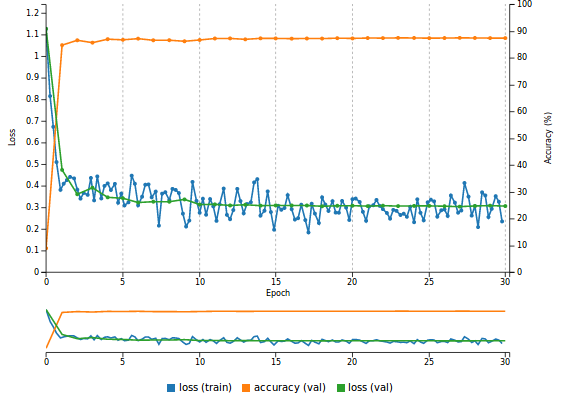}
\caption{The training process of the experiment 3.}
\label{trainprocess_pecu}
\end{center}
\end{figure}

In this case, the images are also distinguished by our strategy. We expected the accuracy of the classification will be good; the results proved this. In Figure~\ref{trainprocess_pecu}, the model obtained an accuracy rate of $87\%$ in the $30$th epoch, and the accuracy rate for each class was also better than that of experiment 2.

\begin{figure}[!htbp]
\graphicspath{{fig/}}
\begin{center}
\includegraphics[scale=0.7]{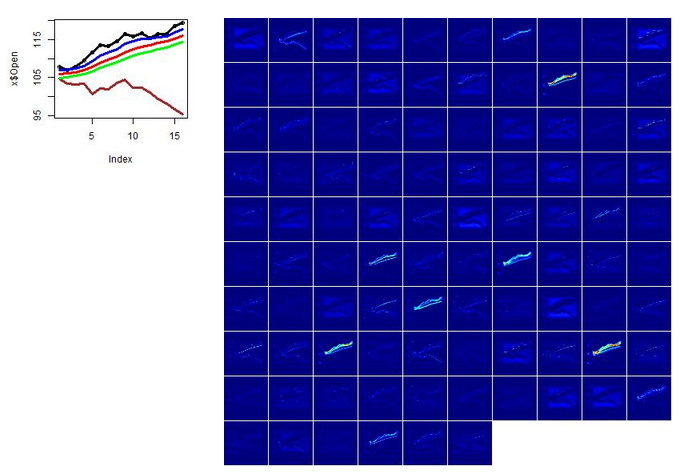}
\caption{The visualization of the first convolution layer with the demo image.}
\label{visual1}
\end{center}
\end{figure}

\begin{figure}[!htbp]
\graphicspath{{fig/}}
\begin{center}
\includegraphics[scale=0.7]{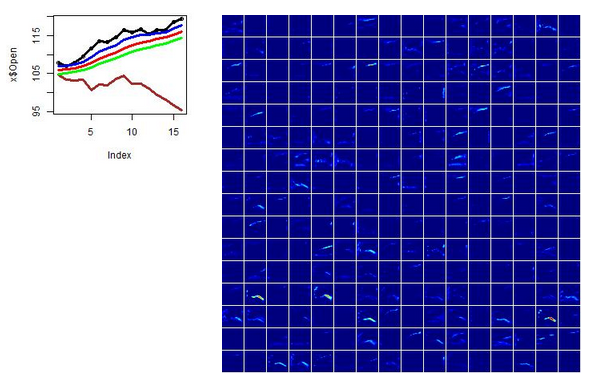}
\caption{The visualization of the second convolution layer with the demo image.}
\label{visual2}
\end{center}
\end{figure}

We also examine the visualization after the convolution layer. The outputs after the first two convolution layers with the demo image are shown in Figures~\ref{visual1} and~\ref{visual2}; we can clearly see that the kernels in the first two layers can capture the shape of the lines. In this image, which is the buy action, the convolution model can clearly capture the pattern of the increasing trend.


\section{Conclusions}
In workflow 1, neither the simple nor complex CNN architectures produced the expected performance. The main cause of this is the lack of data for each convolution model. Attempting to use additional, older historical data would only introduce additional noise and further mislead the convolution model. Therefore, we narrowed the scope of our research to fit in the simulated world, which is generated by applying the GBM calibrated from the real-world data.

In workflow 2, the main difference between the first two experiments (experiment 1 and experiment 2) and the last two experiments (experiment 3 and experiment 4) is the strategies employed. In the first two experiments, the trend of the different labels was not obvious, whereas in the last two experiments, the trend was clearly seen by the human eye. Therefore, the convolution model showcases better performance for the last two strategies, especially for the buy and sell actions.

We conclude that if the strategy is clear enough to make the images obviously distinguishable, then the CNN model can predict the prices of a financial asset; the default AlexNet model is also considered good enough for prediction.

There are additional factors we intend to research in future; for example, combining the convolution model with the other architectures, like the LSTM. The architecture of the time-series model may help the convolution model to capture more information from the pixel images.

\newpage

\bibliographystyle{abbrv}
\bibliography{scibib}

\begin{thebibliography}{1}

\bibitem{browne1995optimal}
S.~Browne.
\newblock Optimal investment policies for a firm with a random risk process:
  Exponential utility and minimizing the probability of ruin.
\newblock {\em Mathematics of Operations Research}, 20(4):937--958, 1995.

\bibitem{diartificial}
L.~Di~Persio and O.~Honchar.
\newblock Artificial neural networks approach to the forecast of stock market
  price movements.
\newblock {\em International Journal of Economics and Management Systems},
  1:158--162, 2016.

\bibitem{fukushima1982neocognitron}
K.~Fukushima and S.~Miyake.
\newblock Neocognitron: A self-organizing neural network model for a mechanism
  of visual pattern recognition.
\newblock In {\em Competition and cooperation in neural nets}, pages 267--285.
  Springer, 1982.

\bibitem{shrevestochastic}
S.~E. Shreve.
\newblock {\em Stochastic Calculus for Finance II: Continuous-Time Models}.
\newblock Springer, New York, 2004.

\bibitem{wang2017origin}
H.~Wang, B.~Raj, and E.~P. Xing.
\newblock On the origin of deep learning.
\newblock {\em arXiv preprint arXiv:1702.07800}, 2017.

\end{thebibliography}

\end{document}